\documentclass[a4paper,10pt,twocolumn,twoside]{article}
\pdfoutput=1
\usepackage{amsmath, amsthm, amssymb}
\usepackage[english]{babel}
\usepackage{chicago}
\usepackage[pdftex]{graphicx}
\usepackage{babelbib} 
\usepackage{hyperref}
\usepackage{authblk}
\usepackage[utf8]{inputenc}
\usepackage{geometry}
\hypersetup{
pdfauthor=Havemann and Scharnhorst,
pdftitle=Bibliometric Networks,
colorlinks=true,
linkcolor=blue,
anchorcolor=blue,
citecolor=blue,
urlcolor=black,
menucolor=black
}

\title{Bibliometric Networks}
\date{}

\author[*]{Frank Havemann}
\author[**]{Andrea Scharnhorst}
\affil[*]{Institut f\"ur Bibliotheks- und Informationswissenschaft der Humboldt-Universit\"at zu Berlin}
\affil[**]{Data Archiving and Networked Services (DANS), Den Haag}

\begin{document}

\maketitle
\begin{abstract}
This text is based on a translation  of a chapter in a handbook about network analysis (published in German) where we tried to make beginners familiar with some basic notions and recent developments of network analysis applied to bibliometric issues~\cite{Havemann2010Bibliometrische}. 
We have added some recent references.
\end{abstract}

\tableofcontents

\section{Introduction}
Bibliometrics is a research field that deals with the statistical analysis of bibliographic information. Most bibliometric research can be categorized as scientometric, because it relates to scientific publication output, in particular to journal articles. Informetrics captures flows of information not just within, but also beyond the world of books and periodicals, including communication over the Web (Webometrics) and the Internet.\footnote{For an introduction to these overlapping fields, we refer the reader to the \textit{Introduction to Informetrics}, a textbook by \citeN{Egghe1990introduction}, to the \textit{Handbook of Quantitative Science and Technology Research}, edited by \citeN{moed2004hqs}, to \textit{Bibliometrics and Citation Analysis}, the most recent English language monograph by \citeN{DeBellis2009Bibliometrics}, and to the recent open-access electronic book by \citeN{Havemann2009Bibliometrie}, \textit{Einführung in die Bibliometrie} (Introduction to Biblometrics, some of the sections in this article are essentially abbreviated versions of sections contained in chapter three of Havemann's book.) Finally, as a good starting point to initiate the subject, we refer the reader as well to the lecture text of Wolfgang \citeN{glanzel2003brf}, \textit{Bibliometrics as a Research Field}.}

The bibliographic description of a written work contains a number of elements such as the name(s) of the author(s), title of the piece, keywords and data necessary for locating the document (e.\,g., title of the journal or edited volume in which an article appears, year of publication, volume/issue number, page numbers). This information is often collected and archived in databases. All of these elements constitute the bibliographic attributes of a document, also called the metadata.

A document's attributes are connected to one another through the document itself---author(s) to journal, keywords to publication date, etc. These connections of different attributes generate bipartite networks which can be represented as rectangular matrices. Like attributes, such as authors with authors (see section \ref{co-auth} below, on co-authorship networks) or keywords with keywords (see section \ref{vector-space} below, on co-word analysis), can also be coupled to one another; this kind of coupling produces unipartite networks, represented by square, symmetrical matrices.

Scientific publications are characterized by the fact that, as a rule, they contain references to other scientific works. This generates a further network, namely, that of the publications themselve. This kind of self-reference has also been taken up in models of the scientific publication process \shortcite{bruckner1990application,gilbert1997ssa,leydesdorff2001csd,morris2003tlv,lucio2009dynamics}.

After Eugene Garfield's pioneering and seminal work, the \textit{Science Citation Index} (SCI), in the 1960s, the SCI was used not only to navigate information, but it was also applied as a tool in scientific analysis \cite{garfield1955cis,Price1965networks,wouters1999citation}. The growing accessibility of machine-readable data (the SCI was already available on magnet tapes relatively early on) and the appearance of the second big information network, the \textit{World Wide Web} (WWW), made large-scale automated data processing and analysis possible. The WWW---in which pages constitute the network nodes and (hyper-)links the edges---has itself become a popular object of research in network analysis \cite{huberman2001laws}. 

Link analysis also proves fruitful when applied to academic institutions or countries, as was shown by \citeN[2009]{thelwall2004link} \nocite{thelwall2009introduction} and also by \shortciteN{ortega2008maps}. 
The position of links in a network, allows us to draw conclusions about research collaboration and the function of different national scientific systems in international research landscape.

In addition, the Web remains today a medium for accessing data, in part freely, which we can analyze according to their own particular network structures. Consider, for example, the bio-information databases, Ebay, or Facebook. Data-flow networks of this kind have resulted in the evolution of a new specialty within statistical physics, namely, that of \textit{complex networks}~\cite{scharnhorst2003complex,Morris2004crossmaps,pyka2009intro}. Within the multidisciplinary setting of network science, the methods of \textit{social network analysis} (SNA), originally developed for smaller social networks, are combined with statistical analysis and dynamic modeling from physics, with computer science algorithms for data mining and visualization, and with graph theory in mathematics, for the purpose of better grasping, explaining, and mastering existing complex networks in nature and society~\cite{nrc2005network}. In recent years, bibliometrics and scientometrics have been strongly influenced by these developments in new network research \shortcite{borner2007network}. We will examine this more closely in section \ref{outlook} below. In what follows immediately, however, we confine ourselves to the network descriptions that have traditionally received special attention in scientometrics.

In scientometrics we distinguish roughly between the analysis of texts and the analysis of actors. Even if information on both elements is contained in a single bibliographic record, historically speaking, most bibliometric network analyses have concentrated on text elements (viz., citation, co-citation, bibliographic coupling, or semantic networks). This stands in contrast to the analysis of scientific collaboration networks---either for cooperation on an individual level or cooperation between countries \cite{wagner2008new}. Among the more interesting methods that have evolved in the intermediate field of text and actor is the HistCite method developed by Eugene Garfield et al., which we discuss in section~\ref{ArtcitationNetworks} below.

\section{Citation Networks \\of Articles}
\label{ArtcitationNetworks} 
\begin{figure}[!b]
 \begin{center}
   \includegraphics[height=7.2cm,width=7.2cm] {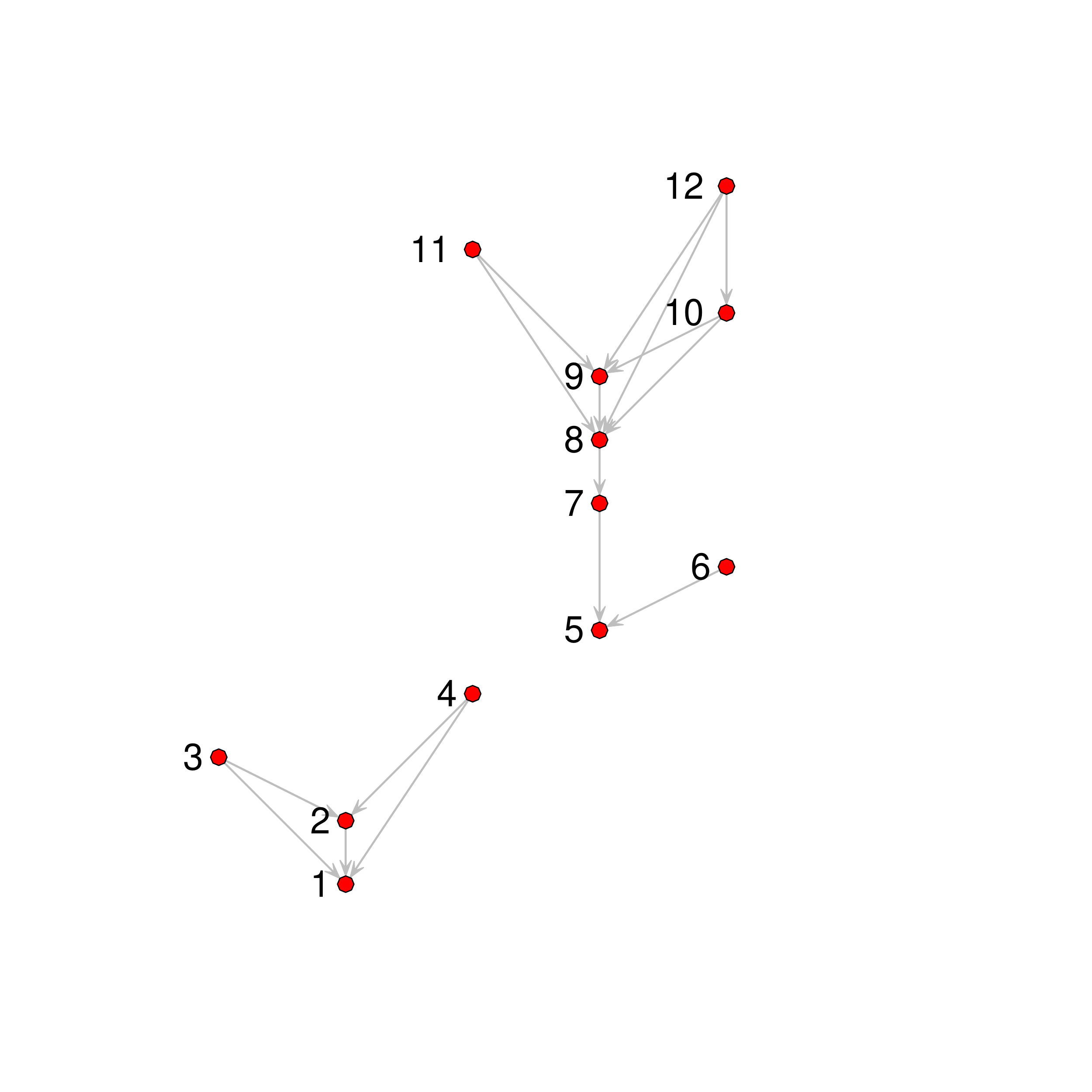}
  \end{center}
\caption{Temporally ordered graph of a citation network for the first twelve articles on N-rays. Data source: de Solla Price (1965)}
\label{fig:N-rays}
\end{figure}
\label{model:reader}

Articles in scientific periodicals base on other knowledge and acknowledge this by referencing earlier articles and other publications. 
This way they build networks. This view of a stream of periodicals literature was already propagated more than 40 years ago by Derek J.~\citeN{Price1965networks}. Defining articles as nodes or vertices and citations as the links or edges of a network graph allows us to apply graph theoretical methods to bibliometrics. In view of the social nature of the science system, we will also consider how terms and concepts initially developed for SNA can be fruitfully applied to explain scientific communication.

New nodes with directed edges pointing to previously available nodes are continually added to both the network of journal articles and the Web. 
Whereas web pages can be modified (or even completely deleted), a journal article remains an unchanged document once it has been published. Corrections, for example, can only be added to the document subsequently as \textit{errata} or \textit{corrigenda}. Hyperlinks can be added to existing pages retroactively, referring to pages that have been subsequently produced. In citation networks, by contrast, there is temporal ordering; but the order, as it were, is not strict. Because of the protractedness of the publication process, authors may be aware of not-yet-published works and cite these, but citation network analysis usually does not take this into account.

Today's vast network of scientific journal articles began to build up when the referencing of older publications became common practice. It might be helpful to try to  visualize the spacial expansion of the whole article citation network as a continually growing sphere which adds a new ``growth ring'' every year, in which articles are located through the sources that they cite. The HistCite method developed by Eugene \shortciteN{garfield2003wna} assumes that, at the root of every citation network, there is a single piece of path-breaking research---namely, the major work of one scientist or the core works emanating from a specialist group or specialty area. Successful strands of research can thus be extracted or distilled from this work. In easily accessible networks of frequently cited articles, the paths of scientific insight and knowledge are clearly visible. This method can also be used to show connections between different scientific schools and communities or their relative isolation from each other \shortcite{lucio2012mathematical}.

As \citeN{lucioarias2008mpa} show, a main path analysis of these temporally ordered graphs reveals the mechanisms of dissemination and diversification (diffusion), as well as those of consolidation and standardization (codification) of scientific knowledge. This kind of citation-based historiography complements biographic and scientific-historic investigation and, in so doing, bridges the graph-theoretical concept of networks and the role of social networks in structuralist social theory \cite{merton1957social}.

If we determine the actual location of a publication only on the basis of information given in the sources cited, then essentially we forgo other information in the document, which may be equally crucial for this determination. Therefore, a reconstruction of knowledge transfer should not buttress itself solely on the analysis of citation networks. Citation analysis does, however, have the advantage of being able to use the mathematical calculations of SNA and thereby achieves results that hermeneutic historiography alone cannot. Moreover the standard methods of information retrieval use the textual similarities between documents to show users of specifically retrieved texts further texts that might be relevant for them (see section ~\ref{vector-space}, S.~\pageref{vector-space}). More recent approaches to information retrieval embrace also other bibliometric regularities \shortcite{mutschke2011science}, co-author patterns, and journal distributions.

Relations (edges, links) in a network of nodes can be captured mathematically in a square, adjacency matrix $A$ whose elements $a_{ij}\ge0$ are different from zero, if node $i$ is related to node~$j$. If we do not differentiate between relations of different strengths, then $a_{ij}$ has a value of 1, if a relationship exists.

In his above-mentioned article, Derek J.\ \citeN{Price1965networks} presented the adjacency matrix for citations between articles in a self-contained bibliography on N-rays, whereby the ones were symbolized with dots and the places that would have been filled by the zeros were left blank \cite[p.\ 514: figure 6]{Price1965networks}.\footnote{N-rays turned out to be fictive; so, for that reason, the bibliography qualifies as self-contained. This concrete example of a citation graph will accompany us through the subsequent sections of this paper.}  He ordered the articles in temporal sequence according to their date of publication and omitted citations of all sources external to the bibliography. The adjacency matrix for the first twelve articles is thus  

\begin{footnotesize}
\begin{eqnarray*}
\label{eq:N-rays}
A =
\left(\begin{array}{rrr rrr rrr rrr}
0 & 0 & 0 & 0 & 0 & 0 & 0 & 0 & 0 & 0 & 0 & 0 \\
1 & 0 & 0 & 0 & 0 & 0 & 0 & 0 & 0 & 0 & 0 & 0 \\
1 & 1 & 0 & 0 & 0 & 0 & 0 & 0 & 0 & 0 & 0 & 0 \\
1 & 1 & 0 & 0 & 0 & 0 & 0 & 0 & 0 & 0 & 0 & 0 \\
0 & 0 & 0 & 0 & 0 & 0 & 0 & 0 & 0 & 0 & 0 & 0 \\
0 & 0 & 0 & 0 & 1 & 0 & 0 & 0 & 0 & 0 & 0 & 0 \\
0 & 0 & 0 & 0 & 1 & 0 & 0 & 0 & 0 & 0 & 0 & 0 \\
0 & 0 & 0 & 0 & 0 & 0 & 1 & 0 & 0 & 0 & 0 & 0 \\
0 & 0 & 0 & 0 & 0 & 0 & 0 & 1 & 0 & 0 & 0 & 0 \\
0 & 0 & 0 & 0 & 0 & 0 & 0 & 1 & 1 & 0 & 0 & 0 \\
0 & 0 & 0 & 0 & 0 & 0 & 0 & 1 & 1 & 0 & 0 & 0 \\
0 & 0 & 0 & 0 & 0 & 0 & 0 & 1 & 1 & 1 & 0 & 0 \\
\end{array}\right).                   
\end{eqnarray*}
\end{footnotesize}

Because future articles could not be cited, ones occur in rows of this matrix only to the left of (or below) the main diagonal. Thus, because of the temporal sequencing of the citation network, de Solla Price's adjacency matrix takes the form of a triangle; along the main diagonal and to the right of (or above) it occur only zeros ($a_{ij}=0, \forall j \ge i$).\footnote{Temporally ordered networks are acyclic. Within acyclic graphs, there is no path along the directed links which loops back to return to the starting point.}

In figure \ref{fig:N-rays}, the graph of the first twelve articles in the citation network disaggregate into two partial graphs. Each of these graphs represents independently achieved results which could only, at a subsequent point in time---namely, with the appearance of the first overview article on N-rays (number 75 in the bibliography)---be interpreted as belonging together.\footnote{This expresses itself as co-citation (cf.\ section~\ref{cocitation}, p.~\pageref{cocitation}).} The adjacency matrix, $A$, of a citation network can be used to model a reader's behavior moving from article to article by following the cited sources. For example, a reader may begin with article 12; the starting time thus noted is $t = 0$. He/she can be described by the column vector,  $\vec r(0)$, which contains eleven zeros and one one as the twelfth component. By multiplying this vector from the left with the transpose of $A$, one finds the reader by articles 8, 9, and 10. In the next step, by means of the rule $\vec r \gets A^{\mathrm{T}}\vec r$, he/she wanders from there to articles 7, 8, and 9, and so forth:

\begin{footnotesize}$$
\vec r(0)=
\left(\begin{array}{r}
0\\
0\\
0\\
0\\
0\\
0\\
0\\
0\\
0\\
0\\
0\\
1\\
\end{array}\right), \vec r(1)=
\left(\begin{array}{r}
0\\
0\\
0\\
0\\
0\\
0\\
0\\
1\\
1\\
1\\
0\\
0\\
\end{array}\right), \vec r(2)=
\left(\begin{array}{r}
0\\
0\\
0\\
0\\
0\\
0\\
1\\
2\\
1\\
0\\
0\\
0\\
\end{array}\right), \ldots
$$
\end{footnotesize}

We can refine this model to one of the random reader,\footnote{This refers to the random surfer model upon which \citeN{Brin1998anatomy} based their PageRank algorithm.} whereby each component of the vector scales to one:
$\vec R = \vec r/\sum r_{i}\equiv  \vec r / r_{+}$.
Thus, for a given time $t = 2$, for example, the components of the reader vector different from zero will be $R_{7}(2)=1/4$, $R_{8}(2)=1/2$,  $R_{9}(2)=1/4$. By scaling to one, we can interpret these fractions as a probability, namely, the probability that, at time $t = 2$ we would encounter the reader by article 7, 8, or 9, should he/she, upon completion of his/her reading, randomly select one of the sources cited. We have a double chance of finding the reader by article 8 because he/she can arrive at 8 via 9 as well as 10. This makes it clear why scaling to one allows us now to designate the reader as a random reader.

Today, with the aid of citation indexes, readers are not only able to search for articles in the reference lists of cited sources retrospectively, but they can also navigate temporally forward in citation networks. 
We can model this process by using the transpose of the transposed adjacency matrix that is, $A$ itself, $(A^{\mathrm{T}})^{\mathrm{T}}=A$, because reflection along the main diagonals reverses all of the arrows in the graph of a directed network, which is easy to show.

The adjacency matrix, $A$, gives the direct paths between the nodes in the network along the directed edges. For the model of the reader, we used powers of $A$ (or $A^{\mathrm{T}}$): $A^{1}, A^{2}, A^{3}\ldots$. For example, if we compare $A^{2}$ with the graph in figure \ref{fig:N-rays}, we see that the components of $A^{2}$ indicate how many (indirect) paths of length 2 there are between the nodes---for instance, there are two such paths between node 8 and node 12 (between the remaining pairs of nodes there is at most one path of length 2). In general, then, a matrix $A^{k}$ contains the number of ways of length $k$ between the vertexes.

\section{Bibliographic Coupling}
\label{bibl-coupl}
In accordance with \citeN{Kessler1963bibliographic}, two articles are said to be bibliographically coupled if at least one cited source appears in the bibliographies or reference lists of both articles. If we look for such bibliographic couplings in the graph of the first twelve articles on N-rays (figure~\ref{fig:N-rays}, p.~\pageref{fig:N-rays}), we discover a number of these. Nodes 2 and 3 are bibliographically coupled over node 1, as are nodes 2 and 4. Nodes 3 and 4 are coupled over node 1 and over node 2. Nodes 6 and 7 are coupled over node 5. Nodes 9 through 12 are coupled in pairs over node 8, and nodes 10 to 12 are coupled in pairs additionally over node 9.

If a reader desires to locate a bibliographically coupled article in a citation network, he/she must first move one step along an arrow to the next vertex, and then, from that node, in the opposite direction of the next arrow. As known from graph theory, this process is described by the matrix $B=AA^{\mathrm{T}}$. Its element $b_{ij}$, in accordance with the rules for matrix multiplication, is the scalar product of the row vectors from $A$. Because $A$ is a binary matrix, summation results in the number of matching components of both row vectors, that is, the number of common sources.

Thus, element $b_{ij}$ indicates how many bibliographic couplings exist between articles $i$ and $j$. In other words, $b_{ij}$ gives the number of paths of length 2, via which one moves from $i$ along the arrow and then to $j$ in the opposite direction. The symmetry of the coupled pairs corresponds to that of the matrix $B=B^{\mathrm{T}}$. The main diagonal contains the numbers of bibliographic self-couplings of an article, namely, the numbers of all of their references to other articles in the network.

Citation databases enable users to move temporally forward, backward, and laterally (by zig\-zagging). Thematically similar articles appearing in the same volume of a journal are often temporally so proximate that the earlier article cannot be cited in the later one. But they also reveal their likeness through similar reference lists, that is, through strong bibliographic coupling. As early as the late-1980s, with the old CD-ROM edition of the Science Citation Index, the user was directed from an article which he/she located to the twenty most strongly bibliographically coupled articles via the ''related records`` option.\footnote{This use of bibliographic coupling for information retrieval is still part of the on-line edition of the Science Citation Index, now part of the Web of Knowledge (see \url{http://wokinfo.com/}).} The strength of the coupling of two articles, $i$ and $j$ is defined here simply by the number of references that the articles have in common, as given by the element $b_{ij}$ of matrix~$B$.

In our example of the bibliography on N-rays, we can only include in our analysis the citation relations between articles in the N-ray bibliography, although clearly other sources have been cited in the articles' reference lists, which do not belong to the bibliography. In an alternative approach, we can analyze a complete body of scientific literature of a publication period together with all cited sources. 
Using the SCI we could, for instance, analyze the publications in one specific year. Rather than selecting a thematic excerpt or segment from the citation graph, we then consider all of the journal articles of that year with all of their cited sources, including books, patents, newspaper articles, etc. Accordingly, the resultant matrix is not a square adjacency matrix $A$ whose rows and columns represent the same vertexes, 
but rather a rectangular matrix, each of whose rows represents an article and each of whose columns represents a cited source. Only a few of the journal articles for that particular year will appear as a source. Thus we arrive at citation network consisting of just two types of nodes: articles and sources. Note that the articles carry the same publication year, and the sources can be from any year. Within this network, only edges between nodes of different type are permitted. Networks of this type are also called \textit{bipartite}; the rectangular matrix is termed an affiliation matrix.\footnote{From the Latin \textit{ad-filiare} meaning to adopt as a son.}

For the rectangular affiliation matrix $A$ with $m$ rows (articles) and $n$ columns (sources), we can also calculate the matrix of bibliographic couplings $B=AA^{\mathrm{T}}$. Matrix $B$ is also square, in this case, and contains for each of the $m$ articles one row and one column.

Two articles, both with long reference lists, could contain many sources in common, in which case they would be said to be strongly bibliographically coupled. Articles with only a few references, therefore, would tend to be more weakly bibliographically coupled, if coupling strength is measured simply according to the number of references articles contain in common. This suggests that it might be more practicable to switch to a relative measure of bibliographic coupling, which we can define most simply by using set theory. In references lists, each cited source appears only once; thus we can see a reference list as a set of cited sources. In set theory language we formulate this thus:
$$b_{ij}=|\mathbf{R}_{i} \cap \mathbf{R}_{j}|,$$
that is, the element $b_{ij}$ of matrix $B$ equals the size of the intersection of the reference lists in articles $i$ and $j$. The Jaccard index (or Jaccard similarity coefficient) gives us a relative measure of the overlap of two sets:\footnote{The Swiss botanist and plant physiologist, Paul Jaccard (1868--1944), defined this index in 1901.}
\begin{equation}
J_{ij}=\frac{|\mathbf{R}_{i} \cap \mathbf{R}_{j}|}{|\mathbf{R}_{i} \cup \mathbf{R}_{j}|}.
\label{def:Jaccard} 
\end{equation}
The Jaccard index of bibliographic coupling is zero if the intersection of the reference lists is empty; it reaches a maximum of one if both lists are identical (because, in this case, the intersection would be equal to the union).

Another relative measure of similarity between sets, which we can use here, is the so-called Salton index:\footnote{The computer scientist Gerard Salton (1927--1995), who lived and worked in the United States, was one of the pioneers in the area of information retrieval (cf.\ Wikipedia). This index was introduced in \citeN{Salton1983imi}. In section 3.6 of the above-cited bibliometrics textbook, \citeN{Havemann2009Bibliometrie} shows how Salton and McGill define their index alternatively as the cosine of the angle between the row vectors of matrix $A$.}
\begin{equation}
S_{ij}=\frac{|\mathbf{R}_{i} \cap \mathbf{R}_{j}|}{\sqrt{|\mathbf{R}_{i}||\mathbf{R}_{j}|}}.
\label{def:Salton} 
\end{equation}
In this case, the average size of the sets is related to the geometric mean of the size of both sets. Here, too, the index reaches a maximum of one for identical sets and a minimum of zero for disjoint ones.

\section{Co-citation Networks}
\label{cocitation}
We speak of the co-citation of two articles when both are cited in a third article. Thus, co-citation can be seen as the counterpart of bibliographic coupling. Returning to our example, we can also find a number of instances of co-citation among the first twelve articles on N-rays (figure~\ref{fig:N-rays}): articles 1 and 2 are co-cited twice (in articles 3 and 4); articles 8 and 9 are co-cited three times (in articles 10, 11, and 12, respectively), and article 10 is co-cited once with article 8 and once with article 9 (in article 12).

In the previous section we explained how the bibliographic coupling matrix $B$ can be obtained from the scalar product of the row vectors of the adjacency matrix $A$. Now, we have to construct the scalar product of the column vectors from $A$, in order to calculate the elements for the co-citation matrix $C$: $$c_{ij}=\sum_{k}a_{ki}a_{kj}.$$ 
Since $A$ is a binary matrix, the summation yields the number of common components of the respective column vectors, that is, the number of cases in which articles appear in the same row or reference list. Written compactly, we calculate $C=A^{\mathrm{T}}A$. Thus our model reader moves within the graph, first, in the opposite direction of the arrow and then, in a second step, with the arrow. Like matrix $B$, matrix $C$ is also symmetric. The main diagonal of $C$ contains the number of cases in which an article is co-cited with itself (which is the case for every citation). The number $c_{ii}$ is therefore the number of all citations of article $i$ in other articles in the network.

Most of the elements in the co-citation matrix $C$ of our example are equal to zero. This is so because the content-related connections between both citation strands only became apparent as co-citation of these papers, initially, with the publication of the first overview article on N-rays (number 75 in the N-rays bibliography and thus not visible in figure \ref{fig:N-rays}). 
Co-citation relations, unlike bibliographic couplings, are subject to change. Many content-related connections can or may only be recognized by later authors at some subsequent point in time; or, conversely, can or may be deemed as no longer essential by later authors.

As with the bibliographic coupling of articles, it is equally reasonable for purposes of co-citation analysis to consider all of the journal articles of a year with all of their cited sources.
We will now analyze the bipartite network of articles and sources introduced in the previous section not according to how the articles are coupled over sources, but rather just the opposite: that is, how co-citation in articles connects the sources to one another. We can also calculate matrix $C$ in accordance with the bipartite network, yielding $C=A^{\mathrm{T}}A$.

In the co-citation analysis of two successive volumes of a bibliographic database,
many co-cited sources in the first year's papers
will appear again as co-cited sources in the second. Coupling strength will vary, however; many new co-cited pairs of sources will join the old ones. The reference lists for a given year's papers, practically speaking, are analogous to the results of an opinion survey designed to ascertain which cited sources are currently seen to be related 
to one another.

The principle of co-citation was first applied by Irina \citeN{Marshakova1973system} in Moscow in a study on laser physics. Independently from Marshakova the principle was also propagated by Henry \citeN{Small1973cocitation}, a scientist from the Institute for Scientific Information (ISI) in Philadelphia, founded by Eugene Garfield in 1960. Since the 1970s, the co-citation perspective has been at the core of bibliometric analyses of specialties, fields, scientific schools, or paradigms in the sense of Kuhn---in other words, co-citation has provided the main thrust underlying our understanding of the social and cognitive theoretical structure of science~\cite{DeBellis2009Bibliometrics}.

All bibliometric networks can be visualized or modeled graphically. (We will come back to this in section \ref{outlook} below.) Historically, co-ci\-tation analysis data were used for the mapping of science and for the first Atlas of Science project~\cite{garfield1981introducing}. For such purposes, the network of co-cited sources is calculated or derived from the reference lists of publications from a single year's papers,
whereby only those sources are considered the citation numbers of which
exceed a certain threshold value (e.\,g., a threshold value of~5). These sources were seen by Henry~G.~\citeN{small1978cited} as \emph{concept symbols}. In a network thus constructed, the next step is to try to determine clusters of sources, whose members have a strong relationship to each other, but relate only weakly to sources in other clusters.\footnote{In network analysis, such clusters of nodes are also called \textit{communities}.}

To determine clusters of similar objects in accordance with the above criteria, a set of algorithms was developed. If we wish to apply these algorithms, we first have to decide which measure of co-citation strength between two sources best applies: the absolute number of co-citations or, for instance, one of the two relative measures presented above---the Jaccard or Salton indexes.

Relative measures of co-citation result in a weak coupling strength among frequently cited sources, which are only rarely co-cited. This seems appropriate because many of the citing authors do not see a closer relationship between those cited \emph{concept symbols}. At the ISI, Henry Small first started working with the Jaccard index and later with the Salton index \cite{small1985ctc}. Irina \citeN{Marshakova1973system} did not use a relative measure. Instead she calculated the expected values for co-citation numbers, based upon the independence of both citation processes, and accepted only those numbers that exceeded the expected values significantly.

In order to get from clusters of \emph{concept symbols} to a map, the next step is aggregation. Hereby, we take all of the nodes in one cluster and draw them together into a point. Then, we take all of the links between each set of two clusters and draw these together into a single link of a determined strength that corresponds to the specific factual distance between those clusters. In this way, we create a network of clusters that can be visualized. The technique of multidimensional scaling or MDS has been frequently used to visualize complex networks on a two--dimensional plane. Today, networks are often visualized using force directed placement or FDP. 

Since the clusters of nodes thus produced, in turn, represent the vertexes in a co-citation network; using an analogous clustering procedure, we can now generate clusters of clusters and so on, until all of the cited sources of a given year's papers are united in a single cluster that represents that part of science indexed by the database used.

Co-citation cannot only be applied on articles. We can also inquire, for example, how often authors are cited together in order to model or map the structure of the expert community. Or, we can analyze co-citations of entire journals (see section \ref{Journal-networks}). In neither case, however, can we expect that 
aggregations of papers
represent just one specific subject.  Authors, for instance, usually deal with several themes---sometimes simultaneously---which can also belong to different specialized areas of expertise. Another problem of author co-citation is the growing tendency toward more research collaboration, which becomes visible in the increasing number of authors per article in some fields. The thematic flexibility of  authors leads in fact to a real problem: If we wish to determine or define authors' areas of activity, thematic flexibility turns out to be a crucial factor whenever we seek to trace dynamic processes in science (on this matter, see section \ref{outlook} below).

\section{Citation Networks \\of Journals}
\label{Journal-networks}
Research has been able to expand so much over the centuries only because new areas of  expertise have continually opened up, and the various research tasks have been distributed accordingly among the expert  communities representing these new fields of scientific endeavor. Each community has created its own specialized journal. Alongside to these specialized journals, journals coexist in which research results of general interest are published. Currently, the open-access movement changes the journal landscape profoundly.\footnote{see the Public Knowledge Project \url{http://pkp.sfu.ca/}} Still, we can assume that the foundation of a journal is a response of a communicative need of a scientific community in one field, one country or across several.

But even the most highly specialized journals contain more than just those articles contributed by the experts in their respective fields of research: these periodicals also contain articles from other research areas that could be of interest for any reader of a particular journal at a given time. The result of this is that the literature from one scientific area is not just to be found in the core journals of that area, but rather that it is broadly scattered according to Bradford's law~\cite{Bradford1934sources}. 

Nevertheless, despite the addition of literature external to a core research field and in accordance with the Porphyrian tree of knowledge, articles in one journal should cite the articles of journals from contiguous fields more often than they would those from fields or disciplines further away. The early study by \citeN{gross1927cla}  was based on this plausible assumption, already. They determined the number of citations of other journals in the general chemistry publication \textit{Journal of the American Chemical Society}---for the purpose of providing librarians with data relevant for library journal selection. However, number-of-journal-citations data gathered in this way are not just influenced by thematic proximity or distance, but also simply by the quantity and quality of the articles in the journal referenced. Journals with similar topics compete for the articles that are most important for further research. For that reason, articles submitted to a journal for publication must undergo a qualitative appraisal process, the so-called \textit{peer review}. The bigger a journal's reputation, the more articles it will be offered, and hence the more rigorous its review process is likely to be. 
Thus scientific periodicals differ not only according to area of expertise but also according to reputation. 

In sum, then, the citation flows in a network of scientific journals are influenced by three main factors: thematic contiguity, the size of a journal, and the reputation of a journal. As a further influencing variable can be added the usual number of cited sources per article for the particular area of specialty in question.

We can correct for journal size by relating the number of citations to the corresponding number of articles available to be cited, as \citeN{Garfield1963new-factors} did when they introduced the \textit{journal impact factor} (JIF). In order to take account of the different citation behavior customary in different fields, \citeN{Pinski1976citation} suggested that instead of relating the number of journal citations to the number of citable articles, this number should be related to the total number of references in all of the cited journal's articles. This is tantamount to a kind of import-export relationship that would also take into account that review articles with long reference lists are on average cited more frequently than original articles publishing research results. Thus journal citation networks, constructed with such a normalization, will just mirror the actual thematic relationships of those periodicals and their respective reputations.

Compared to article citation networks, journals networks will naturally have substantially fewer nodes and are, for that reason, not only more transparent, but also lend themselves more easily to numeric analysis. The citation numbers necessary to construct a journals network are presented in aggregated form in the \textit{Journal Citation Reports} of the Science Citation Index and the Social Sciences Citation Index. In the following, we present two examples of journal network analyses.

\subsection{Citation Flows \\Between Journals}

First, we will examine different variants of a network consisting of five information science journals: (1) \textit{Information Processing and Management}, (2) \textit{Journal of the American Society for Information Science and Technology}, (3) \textit{Journal of Documentation}, (4) \textit{Journal of Information Science}, and (5) \textit{Scientometrics}. 

We begin by constructing a network that has been weighted with reciprocal citation numbers~(including self-citations by the journals in the network). With data from the Social Sciences Edition of the Journal Citation Reports, we derive the following adjacency matrix for the citation window 2006 and the publication window 2002--2006:
\begin{equation}
 A = 
\left(\begin{array}{rrrrr}
 79 &  65  & 15  &  6 &  24 \\
 42 & 182  & 11  & 15 &  44 \\
  6 &  22  & 37  &  8 &   6 \\
 20 &  26  & 13  & 30 &  11 \\
  7 &  48  &  7  & 10 & 254 \\
\end{array}\right).
\label{5-journals-raw}
\end{equation}
Matrix $A$ contains only elements $a_{ij}\neq 0$, because all five journals cite each other as well as themselves. The main diagonal contains the journal self-citations. In the graph of $A$, edges (links) between all the five vertices flow in both directions. Loops represent self-citations.

Let us, now, like \citeN{Pinski1976citation} switch to a network where the number of citations  $a_{ij}$ of journal $j$ by journal $i$ are divided by the sum $a_{j+}$ of the number of references to $j$ in the 5-journals network, yielding $\gamma_{ij}=a_{ij}/a_{j+}$.\footnote{The actual data can be found in the book by \citeN{Havemann2009Bibliometrie}.} 

Pinski and Narin go even further. They argue that citation in a prestigious journal should count more than citation in a less important one. But precisely because they measure prestige itself by number of citations (per cited source), they end up with a recursive concept of that notion, like the concept of prestige that has been debated since the 1940s for social network analysis~\cite{Wasserman1994social}. In social networking, it is not only important how many people one knows; one must know the ``right'' people, namely, those who know a lot of others who are the ``right'' ones.

How do we conceptualize this recursive notion of prestige mathematically? To this end, we will examine a model of prestige redistribution for the five information science journals under consideration here. To begin ($t = 0$), all five journals should have the same weight, so we fix this at~1. The weights are written as a column vector. Analogous to our procedure for the reader model in an article citation network, we then multiply the column vector from the left with the transpose of the adjacency matrix $\vec w(1)=\gamma^{\mathrm{T}}\vec w(0)$. In this way, weights are redistributed within the network. The new column vector contains exactly the row sums of $\gamma^{\mathrm{T}}$, i.\,e.\ $w_{j}(1)=\gamma_{+j}=a_{+j}/a_{j+}$, the import-export ratios of journals. By repeating this procedure many times over, we can see that the weights iteratively approach fixed limit values. In accordance with this, for $t\to\infty$, the following equation holds:\footnote{That this relationship holds not just in our special case, but in every case, is guaranteed by the theory of eigenvalues. Matrices of type $\gamma^{\mathrm{T}}$ have a principal (maximal) eigenvalue of 1, and the iteration determines the corresponding eigenvector.}
$$\vec w = \gamma^{\mathrm{T}}\vec w.$$
This means that the weights determined by the iteration fulfill an equation which can be seen as an expression of the recursive definition of prestige, viz., that the weight of journal $j$ results from the citation relations to all of the other journals (as well as to itself), according to how close these relations are, factored by the weight of the other journals. \citeANP{Pinski1976citation} call this \textit{influence weight}. Determination equations of this type are also referred to as \textit{bootstrap relations}.

Important to note here is that the iteration procedure is somewhat incorrectly labeled ``redistribution.'' The sum of the five weights after the first iteration step is $4.76 < n = 5$. In other words, weight is lost (because the import-export relations of the larger journals are more propitious than those of the smaller ones). However, this discrepancy can be corrected through scaling; for $t\to\infty$, we obtain the normalized components 0.76, 1.33, 1.03, 0.64, and 1.25. By scaling to $n$, it becomes patently clear who the winners and losers of redistribution are.

Following \citeANP{Pinski1976citation}, Nancy \citeN{Geller1978citation} considered a kind of modified redistribution of journal weights. Her starting point was the theory of Markov chains, a special class of stochastic or random processes in which (just as it is for our case here) the situation at time $t + 1$ is completely determined by the situation at time $t$. Instead of using $\gamma$, she took a somewhat differently scaled matrix $\gamma^*$ with the elements $\gamma_{ij}^{*}=a_{ij}/a_{i+}$, whose transpose belongs to the stochastic matrices. Stochastic matrices have the property that they leave  the sum of vector elements unchanged. Such matrices describe true  redistribution; they are therefore well-suited for the description of random processes in which probability is redistributed over possible system states. Nancy \citeANP{Geller1978citation}'s algorithm is also interesting because there are only a few steps that separate it from Google's \textit{PageRank} algorithm as presented in  the textbook by \citeN[section 3.3]{Havemann2009Bibliometrie}. 
This is an example for a method developed for citation networks which became useful also for Web analysis.

\subsection{Citation Environments \\of Individual Journals}

If we consider citation matrices for large groups of journals, we see that many cells in such a matrix are empty and that citation tends to be confined to smaller more densely networked groups~\cite{leydesdorff2007vci}. This is not surprising; it mirrors the real-world design and relations of scientific specialty areas and disciplines. Nevertheless, the delineation of areas remains a problem which has not been satisfactorily resolved: the borders are fluid.

\citeN{leydesdorff2007vci} suggested another method for determining the position of individual journals within the mass of and relative to all of the areas of science. The starting point is an ego network; let us take, for example, the journal \textit{Social Networks}. \textit{Social Networks} has two citation environments. The first consists of those journals, in a specific time frame, that cite articles in \textit{Social Networks} from a unique time period. We could call this group or set of journals the awareness area, spillover area, or influence area of \textit{Social Networks}---in other words, its \textit{citation impact environment.} The second environment consists of those journals that are cited in articles in \textit{Social Networks}---that is, its \textit{knowledge base}. For each group of journals, then, it is possible to carry out the following analysis independently. We ascertain all of the reciprocal citation links for each respective group with the aid of the Journal Citation Reports. \textit{Social Networks}, our starting point, is a member of both environments. For these citation environments we obtain asymmetric matrices like in equation~\ref{5-journals-raw}~(p.\ \pageref{5-journals-raw}).

By applying the process described above, we obtain groups of journals that have similar citation behaviors; these groups can thus be interpreted or defined as specialty areas. The position of the journal whose ego network was at the starting point gives us information about aspects of interdisclipinarity (for example, by applying betweenness centrality) and a possible interface function~\cite{leydesdorff2007vci}. If we repeat this analysis over a sequence of years, the sometimes changing function of a journal in an equally dynamic journal environment becomes visible. In the case of \textit{Social Networks}, what was revealed was that this journal's functioning as a possible bridge between traditional areas of social science and new methods and approaches from network theory in physics could be reduced to a single year, namely, 2004~\cite{leydesdorff2008daj}.

\section{Lexical Coupling \\and Co-Word Analysis}
\label{vector-space}
For computer-supported information retrieval (IR), documents are characterized by the terms used in them. A manageable (not too big) but nevertheless informative set of such terms, comprised mainly of keywords supplied by authors or indexers, or significant words in a document's title, can serve well for IR. In the extreme, all of the words in a document can be taken into account. What is interesting then is the frequency with which these words occur, not counting stop words like ``the,'' ``and,'' ``of,'' and others. The appearance of terms in a collection or set of documents is described by the term-document matrix $A$, whose element $a_{ij}$ tells us how often in document $i$ the term $j$ occurs.

The term-document matrix, like the matrix introduced above consisting of documents and cited sources (see section \ref{bibl-coupl}), describes a bipartite network, namely, one of terms and documents. In this case, however, by taking into account the frequency with which terms occur in a document, we weight the network.

The so-called vector-space model of IR not only reveals the similarity between documents based on the terms used in them (this is lexical coupling), it also shows the relationships between the terms based on their common usage in the documents (this is the basis of co-word analysis). The former corresponds to bibliographic coupling of articles; the latter refers to the co-citation of sources (section \ref{cocitation}).

At the beginning of the 1980s Michel Callon and his group in France proposed co-word analyses as an alternative or supplement to prevailing co-citation methods~\cite{callon1983translations}. So-called ``cognitive scientometrics'' is supposed to make the relationships between documents clearly evident where, for whatever reason, reciprocal citation has occurred only sparsely. Anthony van Raan's group in the Netherlands developed interactive tools for co-word based science maps in the context of evaluations. These maps can make the activities of institutions or countries in specific fields of science visible~\cite{noyons2004science}. Clusters in these networks were interpreted as themes or topics, and temporally hierarchically branched trees provided some insight into the dynamics of scientific areas \cite{rip1984co-word}.

Co-word analyses can be understood as the empirical method of the so-called actor-network theory, a social theory in the field of science and technology studies \cite{latour2005reassembling,DeBellis2009Bibliometrics}. Despite more recent and promising text-based network analyses for identifying innovation-relevant scientific research---some researchers even speak of literature-based discoveries \cite{swanson1986fish,kostoff2008literature}---expert knowledge for the interpretation of text-based agglomerations is still necessary. And, despite decades-long efforts by many groups, as Howard White put it succinctly in a 2007 discussion, we are still not able to analyze and visualize the development and change of scientific paradigms or scientific controversies in such a way that they are accessible to non-experts.\footnote{Howard White, 11th International Conference of Scientometrics and Informetrics, Madrid 2007, workshop on mapping, personal notes.}

This deficit may be due in part to the ambivalence of language. In a study by \citeN{leydesdorff2006measuring}, the authors pointed to the significance of context for words, and they suggested returning to the text-document matrices for word analyses as well, in order to gain more complete information. Probably the answer also lies in the clever selection of a base unit for statistical procedures. An analysis of developing discussion focal points in online forums has shown that, already in one single post, contributors brought up or referred to different topics. Therefore, in this particular case, the choice of sentences as the base unit for statistical network analyses produced better results than did the longer text passages of one post~\cite{prabowo2008evolving}.

In text mining, in sources (such as all keywords, all titles, abstracts, or full text) extraction of terms or phrases is performed according to different algorithms; and statistical measures for frequency and correlation (including network indicators) are applied for which the reference unit such as a phrase, a sentence, a document is an important parameter. The plethora of combinations of these elements presents a great challenge to text analysis and text mining---a challenge which can only be addressed through strong networking of all text-based structural searches including semantic Web research~\cite{vanderEijk2004constructing}.

One method of information retrieval developed for extracting topics from corpora, which uses both types of links in bipartite networks of documents and terms---namely, co-word analysis and lexical coupling---is  latent semantic analysis (LSA) proposed by \citeN{deerwester1990ils}. This method is based on singular value decomposition (SVD) of the term-document matrix. By means of SVD, bipartite networks of articles and cited sources can also be analyzed thematically~\cite{janssens2008hmi,Mitesser2008mdr}.

\section{Co-authorship Networks}
\label{co-auth}
Co-authorship is considered an indicator of cooperation. If two or more authors share the responsibility for a publication presenting particular research results; then, in the course of the research process that led to those results, these individuals ought to have worked together in some way or another at one time. 

It is appropriate and important to agree on a concept of cooperation before the structure of co-authorship networks is analyzed or interpreted~\cite{sonnenwald2007scientific}. To discuss this in-depth, however, would exceed the scope of this article, so we refer the reader to the definition of research collaboration in the aforementioned bibliometrics textbook~\cite[section 3.7]{Havemann2009Bibliometrie} where the author relies on work of Grit~\citeN[S. 32--40]{laudel1999ife}, see also \citeN{laudel2002wdwm}.

Not every form of collaboration finds its ultimate expression in a co-authored work. 
On the other hand, a certain tendency to arbitrarily assign co-authorship to individuals (or force one on them) cannot be denied. 
On the other hand, shared responsibility for an article in a renowned journal is only rarely possible without some form of cooperation. The co-authors, one would expect, at least know each other,\footnote{The likely exception here being articles with 100 or more authors.} and this realistic expectation is what makes the analysis of co-authorship networks interesting.

Co-authorship networks are usually introduced as networks of authors. In its simplest form the co-authorship network is unweighted. A link between two authors exists, if both appear together as authors of at least one publication in the bibliography or body of literature under investigation. Weighting the link with the number of articles in which both  appear together as authors suggests itself, but this kind of modeling still uses only part of the information about cooperation that can actually be extracted from a bibliography. What it fails to capture is whether the relationships between authors are purely bilateral or whether these researchers work together in larger groups. If, for example, three authors cooperate on one article, then between them in total three links of weight 1 can be found; this is exactly the same as if three pairs of them had each published one article together (in total, then, three articles). 

Another aspect which could be taken into account is the sequence in which the co-authors are listed. Though, in different communities there are different rules whom to place first and last, it has been proposed to include information on author sequence in bibliometric indicators \cite{galam2011tailor}.
More complete information is used if co-authorship is presented as a bipartite network of authors and articles. Element $a_{ij}$ of affiliation matrix $A$ is equal to 1 if author $i$ appears among the authors listed for article $j$; otherwise it is equal to zero.

Up to now, most investigations have confined themselves to co-authorship networks in which only authors are represented as nodes and publications are not. The adjacency matrix $B$ of such a network can be calculated from the affiliation matrix $A$, whereby $B=AA^{\mathrm{T}}$. This becomes immediately apparent if we carry our deliberations on networks of bibliographically coupled articles (section \ref{bibl-coupl}) over to the authors-articles network. In a bipartite network, a co-author is someone whom we reach, whenever we take a step toward a publication ($A^{\mathrm{T}}$) and then back again to an author ($A$). We derive the co-authorship figures for two authors from the scalar product of the (binary) row vectors of $A$. The diagonal element $b_{ii}$ in matrix $B$ is therefore equal to the number of publications to which author $i$ was a contributor.

So, how are co-authorship networks structured? First of all, we frequently find that an overwhelming majority of specialty area authors (more than 80\,\%) are grouped together in a single component of the network, namely, the so-called main component. The remaining authors, conversely, often comprise only small groups of researchers connected to one another (at least indirectly) over co-authorship links. All of the distances occurring between the cooperation partners---whether these are functional (subject-related), institutional, geographical, language-related, cultural, or political---cannot prevent the emergence of one big interrelated network of cooperating scientists.

Equally worthy of note, in comparison to size, are the very slight distances between authors in the main components of co-authorship networks. In a co-authorship network consisting of more than one million biomedical science authors whose combined output for the period 1995--99 was more than two million published articles (verified in Medline), the statistical physicist, Mark E.~J.~\citeN{Newman2001scientific-I} found that over 90\,\% of the authors were grouped together in the main component.\footnote{Because authors in bibliographic databases are not always clearly identifiable, the figures vary 
according to the method of identification used. Newman found that, if he took into account authors' full names, the main component would comprise 1.5 million different authors; however, if he included the surnames and only initials for first name, then this figure shrank to just under 1.1 million individuals. For statistical analysis purposes, this ambiguity is only of minor importance; but it strongly impairs the systematic pursuit of individual research. Newer methods depend on a combination of names, addresses, channels of publication, and citation behavior, in order to automatically and unambiguously ascribe researcher identification.} The second largest component of this network contained only 49 authors. The maximal distance between two authors in the main component (also called the network diameter) is a matter of only 24 steps or hops: that is, on the shortest path between two arbitrary nodes, lie a maximum of 23 other nodes. The average length of all of the shortest paths between nodes in the main component is less than five hops~\cite{Newman2001scientific-II}:
\begin{quotation}
\begin{small}
This ``small world'' effect, first described by Milgram,\footnote{\citeN{milgram1967swp}} is, like the existence of a giant component,\footnote{\citeN{Newman2001scientific-I}} probably a good sign for science; it shows that scientific information---discoveries, experimental results, theories---will not have far to travel through the network of scientific acquaintance to reach the ears of those who can benefit by them.  \cite[p.~3]{Newman2001scientific-II}\end{small}
\end{quotation}  
Newman's statement restricts itself to the informal communication of results between researchers, which so often precedes formal publication. This observation confirms one of the two behavioral principles of science Manfred \citeN{bonitz1991impact} formulated early in the history of scientometrics. They read as: 
\begin{quotation}
\begin{small}
\textit{Holographic principle:}
Scientific information ``so behaves'' that it is eventually stored everywhere.
Scientists ``so behave'' that they gain access to their information
from everywhere.

\textit{Maximum speed principle:}
Scientific information ``so behaves'' that it reaches its destination in
the shortest possible time. Scientists ``so behave'' that they acquire their
information in the shortest possible time.                                          \end{small}
\end{quotation}  
Newman's observation confirms the second of Bonitz' principles. 

The famous ``small world'' experiment referred to above, undertaken by social psychologist Stanley~\citeN{milgram1967swp} for the acquaintance network in the US, resulted in an average distance of six hops.\footnote{Milgram's subjects had the task of sending a letter, via their acquaintances, as close as possible to a recipient unknown to themselves. The letters that actually reached the targeted individual had been forwarded on average by six persons (including the subject).} \citeANP{Newman2001scientific-II} explains the small-world effect taking himself as an example: he has 26 co-authors who, in turn, author publications together with a total of 623 other researchers.
\begin{quotation}
\begin{small}
The ``radius'' of the whole network around me is reached when the number of neighbors within that radius equals the number of scientists in the giant component of the network, and if the increase in numbers of neighbors with distance continues at the impressive rate [\ldots]\ it will not take many steps to reach this point.~\cite[p.~3]{Newman2001scientific-II}
\end{small}
\end{quotation}

The number of co-authors that an author has is a measure of his/her interconnectedness. It is equal to the number of his/her links (degree of the node) in the co-authorship network. In a given specialty area, marginal or peripheral authors have only a few cooperation partners. In network analysis, therefore, the degree of a node is also a measure of its centrality. Often the distribution of co-authorship is skewed: a few authors have many co-authors, many authors have only a few co-authors~\cite[p.~5]{Newman2001scientific-I}.

Another measure of centrality is the betweenness of a given node $i$. Betweenness is defined as the total number of shortest paths between arbitrary pairs of nodes, which run through node~$i$. Conceivable then, is that nodes with higher values
of betweenness are responsible for shorter distances in networks and also for the emergence of large main components. And, in terms of betweenness centrality, a number of frontrunners also set themselves clearly apart from the remaining authors~\cite[p.~2]{Newman2001scientific-II}.

Finally, the different roles and functions of researchers in co-authorship networks is a topic that has  begun to receive increasing attention. \citeN{lambiotte2009communities} have recently considered the importance of a researcher's function (viz., having a high level of connectivity and authority within a community versus being an facilitator or communicator between communities) for the dissemination of new ideas.

\section{Outlook}
\label{outlook}

Paul Otlet, the pioneer in knowledge organization inherited us a wealth of drawings of the evolving universe of knowledge \cite{vanHeuvel2011facing}. Among them, he depicted the main classes of his decimal classification scheme\footnote{Universal Decimal Classification UDC, see also \url{http://udcc.org/}} as longitudes of a globe of knowledge~\cite{heuvel2005visualizing,vandenHeuvel2008building}. Katy B{\"o}rner has inspired an artistic visualization of the sciences as a growing ``organism'' \cite{Boerner2009161}. Recent efforts for a geography or geology of the expanding globe of scientific literature, have led to a revival of ``science maps.'' The Atlas of Science, presented by the ISI in the early 1980s~\cite{garfield1981introducing} has been followed by a New Atlas of Science \cite{boerner2010atlas}.\footnote{The Atlas of Science of Katy B{\"o}rner captures parts of a remarkable long-term project, called ``Places and Spaces'', which is a growing exhibition of science maps. This exhibition is particularly interesting, for one, because the public invitation and selection process enables highly varied and unique depictions to achieve greater visibility through public display and, second, because the themes of the yearly iterations embrace such a broad and diverse spectrum of subject matter---running, for example, from the history of cartography, over maps as deceptive or phantasy pictures, up to maps drawn by children. Many of the maps on display are based on network data. For more information, see \url{http://www.scimaps.org}.} Nevertheless scientists are still struggling to find adequate imagery or, as the case may be, an appropriate mathematical model to represent the deeper understanding we have of the dynamic processes of evolving science networks.

Against this backdrop of the ``new network science'', the future of bibliometric network analyses lies in the incorporation of methods and tools from other disciplinary areas. Visualizations represent a possible platform for interdisciplinary encounters~\cite{boerner2010atlas}. These new ``maps of science'' have met with similar controversy to that which we know from the history of geographical maps. It therefore comes as no surprise that mappers of science have sought to build bridges to cartography. If we apply the methods of structure identification~(self-organized maps) as they were developed for complexity research~\cite{agarwal2008self}, then it is just a small step from the question of possible models to the explication of complex structure formation.

For bibliometric networks, next to the traditional topic of structure identity, the topic of structure formation---and with it, the dimension of time---steps more forcefully into the foreground of interest. All of the methods we have used up to now in the global cartography of science or knowledge landscapes \cite{scharnhorst2001constructing} confirm the existence of collectively generated self-organizing structures in the form of scientific disciplines and scientific communities. On the macro-level, these patterns are so persistent that, as  \citeN{klavans2009toward} have recently shown, they manifest themselves recurrently, relatively independent of the respective research methods.\footnote{The ring structure of the consensus map bears astounding similarity to Otlet's historical visions of a  globe of knowledge.}

It is more difficult to find general patterns or regularities on the micro-level for the interactions between authors or for those between authors and documented knowledge. What remains is a deeper understanding of the dynamic mechanisms that describe the emergence of a science landscape and individual navigation within it. We expect dynamic mathematical models to reproduce already known statistical bibliometric laws like Lotka's law of scientific productivity~\cite{Lotka1926frequency}  or Bradford's  law of scattering~\cite{Bradford1934sources}, mentioned above. Complexity research with notions like energy, entropy, or fitness landscapes~\cite{scharnhorst2001constructing} offers a rich repertoire of contemporary analytical methods and models which can do justice to the network character of complex systems~\cite{fronczak2007analysis}. Recent empirical research in this area is devoted to contemporary effects in evolving bibliometric networks~\cite{Borner2004simultaneous} and the search for burst phenomena~\cite{chen2009towards} or the application of epidemic models to the dissemination of ideas~\cite{bettencourt2009scientific}. But also on the level of conceptual models philosophy and sociology of science have embraced the idea of an epistemic landscape and mathematical models for the search behavior of researcher in it~\cite{payette2012agent,edmonds2011simulating}.

Topic delineation on both the micro as the macro level is a pertinent problem of science studies in general and bibliometrics in particular. However, the problem of topic delineation in citation-based networks of papers might be a consequence of the overlap of thematic structures. The overlap of themes in publications is well known to science studies. A recent study by \citeN{havemann_identifying_2012} on the meso level tests three local approaches to the identification of overlapping communities developed in the last years~\cite{fortunato2010community}.  
 
The heuristic value of dynamic models lies in their broad range of available possible mechanisms, forms of interaction, and feedback, which can be connected to distinctive types of structure formation. Thus, for example, the topically relevant notion of field mobility\footnote{The term ``field mobility'' was introduced in bibliometrics by Jan \citeN{vlachy1978field} to describe the thematic wandering of scientists. Thematic wandering results from new discoveries as well as other grounds such as the connectivity between scientific areas~\cite{bruckner1990application}.}
can be drawn upon for the explication of growth processes in competing scientific areas, for which, in addition, schools of knowledge as sources of self-accelerating growth are also highly relevant~\cite{bruckner1986Bemerkungen}. Through mobility, a network is created between scientific areas, and, at the same time, all of the elementary dynamic model mechanisms and a quasi ``metabolic network'' of knowledge systems are formed \cite{Ebeling2009Selbstorganisation}. The wanderings of the researcher can be followed or read from his/her self-citations. Whereas self-referencing is usually ignored in scientometrics, these citations constitute an excellent source for the investigation and analysis of mobility in scientific fields. Structures in the self-citation network can be interpreted as thematic areas which become visible through different groups of keywords and co-authorships. The wandering of an individual between research areas, as shown by his/her lifework of collected scientific output, can be displayed or represented as a unique bar code pattern \cite{hellsten2007self}. This creative fuzziness of the scientist can also be shown on science maps as a spreading phenomenon, whereby the scientist, instead of the wandering dot, is depicted as a flow field~\cite{skupin2009discrete}.

The use of models as generators of hypotheses presupposes that the hypotheses have been empirically tested and, accordingly, put into the context of social communication, socioeconomic, and political theories of ``science qua social system'' \cite{glaser2006wissenschaftliche}. This connection to traditionally strongly sociologically anchored SNA---especially the proposed inclusion of social, cognitive, and personality attributes of actors for modeling the evolution of networks and dissemination phenomena within networks---and the inclusion/application of dynamic modeling approaches in SNA provide an excellent basic framework for using models to generate theory~\cite{snijders2010introduction}.

Semantic web applications creating new linked repositories of data, publication and concepts; large scale data mining techniques (as originating from Artificial Intelligence), meaningful visualizations, and mathematical models of science all contribute to a better understanding of the science system. However, similar to the variety of possible network visualizations is the variety of mathematical and conceptual models of science. Often knowledge about data mining, modeling and visualizing is inherited by different relative isolated academic tribes~\cite{becher2001academic}. Translating and linking concepts and methods where appropriate and possible is one remaining task. Exploring and better understanding our own science history is another one. Only if we succeed in bridging unique individual science biographies with the laws and regularities of the science system as a whole, will we be able to learn more about the development of new ideas---knowledge generation---and be better able to intervene in this process in a more controlled and supportive manner.

\section*{Acknowledgement}
We thank Mary Elizabeth Kelley-Bibra for helping us with the translation of the text into English.

\bibliography{informetrics}

\begin{thebibliography}{}

\bibitem[\protect\citeauthoryear{Agarwal and Skupin}{Agarwal and
  Skupin}{2008}]{agarwal2008self}
Agarwal, P. and A.~Skupin (2008).
\newblock {\em {Self-organising maps: applications in geographic information
  science}}.
\newblock Wiley-Blackwell.

\bibitem[\protect\citeauthoryear{Becher and Trowler}{Becher and
  Trowler}{2001}]{becher2001academic}
Becher, T. and P.~R. Trowler (2001).
\newblock {\em {Academic Tribes and Territories: Intellectual enquiry and the
  culture of disciplines}\/} (2nd ed.).
\newblock Society for Research into Higher Education / Open University Press.

\bibitem[\protect\citeauthoryear{Bettencourt, Kaiser, and Kaur}{Bettencourt
  et~al.}{2009}]{bettencourt2009scientific}
Bettencourt, L.~M., D.~I. Kaiser, and J.~Kaur (2009).
\newblock Scientific discovery and topological transitions in collaboration
  networks.
\newblock {\em Journal of Informetrics\/}~{\em 3\/}(3), 210--221.
\newblock Special Edition on the Science of Science: Conceptualizations and
  Models of Science, ed. by Katy B{\"o}rner \& Andrea Scharnhorst.

\bibitem[\protect\citeauthoryear{Bonitz}{Bonitz}{1991}]{bonitz1991impact}
Bonitz, M. (1991).
\newblock The impact of behavioral principles on the design of the system of
  scientific communication.
\newblock {\em Scientometrics\/}~{\em 20\/}(1), 107--111.

\bibitem[\protect\citeauthoryear{B{\"o}rner}{B{\"o}rner}{2010}]{boerner2010atlas}
B{\"o}rner, K. (2010).
\newblock {\em {{Atlas of Science: Visualizing What We Know}}}.
\newblock MIT Press.

\bibitem[\protect\citeauthoryear{B{\"o}rner, Maru, and Goldstone}{B{\"o}rner
  et~al.}{2004}]{Borner2004simultaneous}
B{\"o}rner, K., J.~T. Maru, and R.~L. Goldstone (2004).
\newblock The simultaneous evolution of author and paper networks.
\newblock {\em Proceedings of the National Academy of Sciences of the United
  States of America\/}~{\em 101}, 5266--5273.

\bibitem[\protect\citeauthoryear{B{\"o}rner, Sanyal, and Vespignani}{B{\"o}rner
  et~al.}{2007}]{borner2007network}
B{\"o}rner, K., S.~Sanyal, and A.~Vespignani (2007).
\newblock {Network science}.
\newblock {\em Annual Review of Information Science and Technology\/}~{\em 41},
  537--607.

\bibitem[\protect\citeauthoryear{B{\"o}rner and Scharnhorst}{B{\"o}rner and
  Scharnhorst}{2009}]{Boerner2009161}
B{\"o}rner, K. and A.~Scharnhorst (2009).
\newblock Visual conceptualizations and models of science.
\newblock {\em Journal of Informetrics\/}~{\em 3\/}(3), 161--172.
\newblock Science of Science: Conceptualizations and Models of Science.

\bibitem[\protect\citeauthoryear{Bradford}{Bradford}{1934}]{Bradford1934sources}
Bradford, S.~C. (1934).
\newblock Sources of information on specific subjects.
\newblock {\em Engineering\/}~{\em 137}, 85--86.

\bibitem[\protect\citeauthoryear{Brin and Page}{Brin and
  Page}{1998}]{Brin1998anatomy}
Brin, S. and L.~Page (1998).
\newblock The anatomy of a large-scale hypertextual {Web} search engine.
\newblock {\em Computer Networks and ISDN Systems\/}~{\em 30\/}(1--7),
  107--117.

\bibitem[\protect\citeauthoryear{Bruckner, Ebeling, and Scharnhorst}{Bruckner
  et~al.}{1990}]{bruckner1990application}
Bruckner, E., W.~Ebeling, and A.~Scharnhorst (1990).
\newblock {The application of evolution models in scientometrics}.
\newblock {\em Scientometrics\/}~{\em 18\/}(1), 21--41.

\bibitem[\protect\citeauthoryear{Bruckner and Scharnhorst}{Bruckner and
  Scharnhorst}{1986}]{bruckner1986Bemerkungen}
Bruckner, E. and A.~Scharnhorst (1986).
\newblock {Bemerkungen zum Problemkreis einer wissenschafts-wissenschaftlichen
  Interpretation der Fisher-Eigen-Gleichung dargestellt an Hand
  wissenschaftshistorischer Zeugnisse}.
\newblock {\em Wissenschaftliche Zeitschrift der Humboldt-Universit{\"a}t zu
  Berlin. Mathema\-tisch-naturwissenschaftliche Reihe\/}~{\em 35\/}(5),
  481--493.

\bibitem[\protect\citeauthoryear{Callon, Courtial, Turner, and Bauin}{Callon
  et~al.}{1983}]{callon1983translations}
Callon, M., J.~P. Courtial, W.~A. Turner, and S.~Bauin (1983).
\newblock {From translations to problematic networks: An introduction to
  co-word analysis}.
\newblock {\em Social Science Information\/}~{\em 22\/}(2), 191--235.

\bibitem[\protect\citeauthoryear{Chen, Chen, Horowitz, Hou, Liu, and
  Pellegrino}{Chen et~al.}{2009}]{chen2009towards}
Chen, C., Y.~Chen, M.~Horowitz, H.~Hou, Z.~Liu, and D.~Pellegrino (2009).
\newblock {Towards an Explanatory and Computational Theory of Scientific
  Discovery}.
\newblock {\em Journal of Informetrics\/}.
\newblock Special Edition on the Science of Science: Conceptualizations and
  Models of Science, ed. by Katy B{\"o}rner \& Andrea Scharnhorst.

\bibitem[\protect\citeauthoryear{{De Bellis}}{{De
  Bellis}}{2009}]{DeBellis2009Bibliometrics}
{De Bellis}, N. (2009).
\newblock {\em {Bibliometrics and Citation Analysis. From the Science Citation
  Index to Cybermetrics}}.
\newblock Lanham: The Scarecrow Press.
\newblock ISBN-13: 978-0-8108-6713-0.

\bibitem[\protect\citeauthoryear{de~Solla~Price}{de~Solla~Price}{1965}]{Price1965networks}
de~Solla~Price, D.~J. (1965).
\newblock Networks of scientific papers.
\newblock {\em Science\/}~{\em 149}, 510--515.

\bibitem[\protect\citeauthoryear{Deerwester, Dumais, Furnas, Landauer, and
  Harshman}{Deerwester et~al.}{1990}]{deerwester1990ils}
Deerwester, S., S.~T. Dumais, G.~W. Furnas, T.~K. Landauer, and R.~Harshman
  (1990).
\newblock {Indexing by latent semantic analysis}.
\newblock {\em Journal of the American Society for Information Science\/}~{\em
  41\/}(6), 391--407.

\bibitem[\protect\citeauthoryear{Ebeling and Scharnhorst}{Ebeling and
  Scharnhorst}{2009}]{Ebeling2009Selbstorganisation}
Ebeling, W. and A.~Scharnhorst (2009).
\newblock {Selbst\-organisation und Mobilit{\"a}t von Wissenschaftlern --
  Modelle f{\"u}r die Dynamik von Problemfeldern und Wissenschaftsgebieten}.
\newblock In W.~Ebeling and H.~Parthey (Eds.), {\em {Selbstorganisation in
  Wissenschaft und Technik. Wissenschaftsforschung Jahrbuch 2008}}, pp.\
  9--27. Wissenschaftlicher Verlag Berlin.

\bibitem[\protect\citeauthoryear{Edmonds, Gilbert, Ahrweiler, and
  Scharnhorst}{Edmonds et~al.}{2011}]{edmonds2011simulating}
Edmonds, B., N.~Gilbert, P.~Ahrweiler, and A.~Scharnhorst (2011).
\newblock Simulating the social processes of science.
\newblock {\em Journal of Artificial Societies and Social Simulation\/}~{\em
  14\/}(4), 1--6.
\newblock \url{http://jasss.soc.surrey.ac.uk/14/4/14.html} (Special issue).

\bibitem[\protect\citeauthoryear{Egghe and Rousseau}{Egghe and
  Rousseau}{1990}]{Egghe1990introduction}
Egghe, L. and R.~Rousseau (1990).
\newblock {\em Introduction to Informetrics. Quantitative Methods in Library
  and Information Science}.
\newblock Elsevier.

\bibitem[\protect\citeauthoryear{Fortunato}{Fortunato}{2010}]{fortunato2010community}
Fortunato, S. (2010).
\newblock Community detection in graphs.
\newblock {\em Physics Reports\/}~{\em 486\/}(3-5), 75--174.

\bibitem[\protect\citeauthoryear{Fronczak, Fronczak, and Ho{\l}yst}{Fronczak
  et~al.}{2007}]{fronczak2007analysis}
Fronczak, P., A.~Fronczak, and J.~A. Ho{\l}yst (2007).
\newblock {Analysis of scientific productivity using maximum entropy principle
  and fluctuation-dissipation theorem}.
\newblock {\em Physical Review E\/}~{\em 75\/}(2), 26103.

\bibitem[\protect\citeauthoryear{Galam}{Galam}{2011}]{galam2011tailor}
Galam, S. (2011).
\newblock Tailor based allocations for multiple authorship: a fractional
  gh-index.
\newblock {\em Scientometrics\/}~{\em 89\/}(1), 365--379.

\bibitem[\protect\citeauthoryear{Garfield}{Garfield}{1955}]{garfield1955cis}
Garfield, E. (1955).
\newblock {Citation indexes for science. A new dimension in documentation
  through association of ideas}.
\newblock {\em Science\/}~{\em 122\/}(3159), 108--111.

\bibitem[\protect\citeauthoryear{Garfield}{Garfield}{1981}]{garfield1981introducing}
Garfield, E. (1981).
\newblock {Introducing the ISI Atlas of Science: Biochemistry and Molecular
  Biology}.
\newblock {\em Current Contents\/}~{\em 42}, 279--87.
\newblock Reprinted: Essays of an Information Scientist, Vol. 5, p. 279--287,
  1981--82
  \url{http://www.garfield.library.upenn.edu/essays/v5p279y1981-82.pdf}.

\bibitem[\protect\citeauthoryear{Garfield, Pudovkin, and Istomin}{Garfield
  et~al.}{2003}]{garfield2003wna}
Garfield, E., A.~I. Pudovkin, and V.~S. Istomin (2003).
\newblock {Why do we need algorithmic historiography?}
\newblock {\em Journal of the American Society for Information Science and
  Technology\/}~{\em 54\/}(5), 400--412.

\bibitem[\protect\citeauthoryear{Garfield and Sher}{Garfield and
  Sher}{1963}]{Garfield1963new-factors}
Garfield, E. and I.~H. Sher (1963).
\newblock New factors in the evaluation of scientific literature through
  citation indexing.
\newblock {\em American Documentation\/}~{\em 14\/}(3), 195--201.
\newblock \url{http://www. garfield. library. upenn. edu/essays/v6p492y1983.
  pdf}, 2005-4-28.

\bibitem[\protect\citeauthoryear{Geller}{Geller}{1978}]{Geller1978citation}
Geller, N.~L. (1978).
\newblock {On the citation influence methodology of Pinski and Narin}.
\newblock {\em Information Processing \& Management\/}~{\em 14\/}(2), 93--95.

\bibitem[\protect\citeauthoryear{Gilbert}{Gilbert}{1997}]{gilbert1997ssa}
Gilbert, N. (1997).
\newblock {A simulation of the structure of academic science}.

\bibitem[\protect\citeauthoryear{Gl{\"a}nzel}{Gl{\"a}nzel}{2003}]{glanzel2003brf}
Gl{\"a}nzel, W. (2003).
\newblock {Bibliometrics as a research field: A course on theory and
  application of bibliometric indicators}.
\newblock Courses Handout, \url{http://www. norslis. net/2004/Bib_Module_KUL.
  pdf}.

\bibitem[\protect\citeauthoryear{Gl{\"a}ser}{Gl{\"a}ser}{2006}]{glaser2006wissenschaftliche}
Gl{\"a}ser, J. (2006).
\newblock {\em {Wissenschaftliche Produktionsgemeinschaften: Die soziale
  Ordnung der Forschung}}.
\newblock Campus Verlag.

\bibitem[\protect\citeauthoryear{Gross and Gross}{Gross and
  Gross}{1927}]{gross1927cla}
Gross, P. L.~K. and E.~M. Gross (1927).
\newblock {College Libraries and Chemical Education}.
\newblock {\em Science\/}~{\em 66\/}(1713), 385--389.

\bibitem[\protect\citeauthoryear{Havemann}{Havemann}{2009}]{Havemann2009Bibliometrie}
Havemann, F. (2009).
\newblock {\em Einf{\"u}hrung in die Bibliometrie}.
\newblock Berlin: Gesellschaft f{\"u}r Wissenschaftsforschung.
\newblock \\\url{http://d-nb.info/993717780}.

\bibitem[\protect\citeauthoryear{Havemann, Gl{\"a}ser, Heinz, and
  Struck}{Havemann et~al.}{2012}]{havemann_identifying_2012}
Havemann, F., J.~Gl{\"a}ser, M.~Heinz, and A.~Struck (2012, March).
\newblock Identifying overlapping and hierarchical thematic structures in
  networks of scholarly papers: A comparison of three approaches.
\newblock {\em {PLoS} {ONE}\/}~{\em 7\/}(3), e33255.

\bibitem[\protect\citeauthoryear{Havemann and Scharnhorst}{Havemann and
  Scharnhorst}{2010}]{Havemann2010Bibliometrische}
Havemann, F. and A.~Scharnhorst (2010).
\newblock {Bibliometrische Netzwerke}.
\newblock In C.~Stegbauer and R.~H\"au{\ss}ling (Eds.), {\em Handbuch
  Netzwerkforschung}, pp.\  799--823. VS Verlag f\"ur Sozialwissenschaften.
\newblock 10.1007/978-3-531-92575-2\_70.

\bibitem[\protect\citeauthoryear{Hellsten, Lambiotte, Scharnhorst, and
  Ausloos}{Hellsten et~al.}{2007}]{hellsten2007self}
Hellsten, I., R.~Lambiotte, A.~Scharnhorst, and M.~Ausloos (2007).
\newblock {Self-citations, co-authorships and keywords: A new approach to
  scientists’ field mobility?}
\newblock {\em Scientometrics\/}~{\em 72\/}(3), 469--486.

\bibitem[\protect\citeauthoryear{Huberman}{Huberman}{2001}]{huberman2001laws}
Huberman, B.~A. (2001).
\newblock {\em {The laws of the Web: patterns in the ecology of information}}.
\newblock MIT Press.

\bibitem[\protect\citeauthoryear{Janssens, Gl{\"a}nzel, and {De Moor}}{Janssens
  et~al.}{2008}]{janssens2008hmi}
Janssens, F., W.~Gl{\"a}nzel, and B.~{De Moor} (2008).
\newblock {A hybrid mapping of information science}.
\newblock {\em Scientometrics\/}~{\em 75\/}(3), 607--631.

\bibitem[\protect\citeauthoryear{Kessler}{Kessler}{1963}]{Kessler1963bibliographic}
Kessler, M.~M. (1963).
\newblock Bibliographic coupling between scientific papers.
\newblock {\em American Documentation\/}~{\em 14}, 10--25.

\bibitem[\protect\citeauthoryear{Klavans and Boyack}{Klavans and
  Boyack}{2009}]{klavans2009toward}
Klavans, R. and K.~Boyack (2009).
\newblock {Toward a consensus map of science}.
\newblock {\em Technology\/}~{\em 60}, 2.

\bibitem[\protect\citeauthoryear{Kostoff}{Kostoff}{2007}]{kostoff2008literature}
Kostoff, R.~N. (2007).
\newblock {Literature-related discovery (LRD): Introduction and background}.
\newblock {\em Technological Forecasting \& Social Change\/}~{\em 75\/}(2),
  165--185.

\bibitem[\protect\citeauthoryear{Lambiotte and Panzarasa}{Lambiotte and
  Panzarasa}{2009}]{lambiotte2009communities}
Lambiotte, R. and P.~Panzarasa (2009).
\newblock Communities, knowledge creation, and information diffusion.
\newblock {\em Journal of Informetrics\/}~{\em 3\/}(3), 180--190.

\bibitem[\protect\citeauthoryear{Latour}{Latour}{2005}]{latour2005reassembling}
Latour, B. (2005).
\newblock {\em {Reassembling the social: An introduction to
  actor-network-theory}}.
\newblock Oxford University Press, USA.

\bibitem[\protect\citeauthoryear{Laudel}{Laudel}{1999}]{laudel1999ife}
Laudel, G. (1999).
\newblock {Interdisziplin{\"a}re Forschungskooperation: Erfolgsbedingungen der
  Institution 'Sonderforschungsbereich'}.
\newblock {\em Berlin, Edition Sigma\/}.

\bibitem[\protect\citeauthoryear{Laudel}{Laudel}{2002}]{laudel2002wdwm}
Laudel, G. (2002).
\newblock What do we measure by co-authorships?
\newblock {\em Research Evaluation\/}~{\em 11\/}(1), 3--15.

\bibitem[\protect\citeauthoryear{Leydesdorff}{Leydesdorff}{2001}]{leydesdorff2001csd}
Leydesdorff, L. (2001).
\newblock {\em {The Challenge of Scientometrics: the development, measurement,
  and self-organization of scientific communications}}.
\newblock Upublish. Com.

\bibitem[\protect\citeauthoryear{Leydesdorff}{Leydesdorff}{2007}]{leydesdorff2007vci}
Leydesdorff, L. (2007).
\newblock {Visualization of the citation impact environments of scientific
  journals: An online mapping exercise}.
\newblock {\em Journal of the American Society for Information Science and
  Technology\/}~{\em 58\/}(1), 25--38.

\bibitem[\protect\citeauthoryear{Leydesdorff and Hellsten}{Leydesdorff and
  Hellsten}{2006}]{leydesdorff2006measuring}
Leydesdorff, L. and I.~Hellsten (2006).
\newblock {Measuring the meaning of words in contexts: An automated analysis of
  controversies about 'Monarch butterflies,' 'Frankenfoods,' and 'stem cells'}.
\newblock {\em Scientometrics\/}~{\em 67\/}(2), 231--258.

\bibitem[\protect\citeauthoryear{Leydesdorff and Schank}{Leydesdorff and
  Schank}{2008}]{leydesdorff2008daj}
Leydesdorff, L. and T.~Schank (2008).
\newblock Dynamic animations of journal maps: Indicators of structural changes
  and interdisciplinary developments.
\newblock {\em Journal of the American Society for Information Science and
  Technology\/}~{\em 59\/}(11), 1810--1818.

\bibitem[\protect\citeauthoryear{Lotka}{Lotka}{1926}]{Lotka1926frequency}
Lotka, A.~J. (1926).
\newblock The frequency distribution of scientific productivity.
\newblock {\em Journal of the Washington Academy of Sciences\/}~{\em 16\/}(12),
  317--323.

\bibitem[\protect\citeauthoryear{Lucio-Arias and Leydesdorff}{Lucio-Arias and
  Leydesdorff}{2008}]{lucioarias2008mpa}
Lucio-Arias, D. and L.~Leydesdorff (2008).
\newblock {Main-path analysis and path-dependent transitions in HistCite-based
  historiograms}.
\newblock {\em Journal of the American Society for Information Science and
  Technology\/}~{\em 59\/}(12), 1948--1962.

\bibitem[\protect\citeauthoryear{Lucio-Arias and Leydesdorff}{Lucio-Arias and
  Leydesdorff}{2009}]{lucio2009dynamics}
Lucio-Arias, D. and L.~Leydesdorff (2009).
\newblock {The dynamics of exchanges and references among scientific texts, and
  the autopoiesis of discursive knowledge}.
\newblock {\em Journal of Informetrics\/}.

\bibitem[\protect\citeauthoryear{Lucio-Arias and Scharnhorst}{Lucio-Arias and
  Scharnhorst}{2012}]{lucio2012mathematical}
Lucio-Arias, D. and A.~Scharnhorst (2012).
\newblock Mathematical approaches to modeling science from an
  algorithmic-historiography perspective.
\newblock In A.~Scharnhorst, K.~B\"orner, and P.~Besselaar (Eds.), {\em Models
  of Science Dynamics}, Understanding Complex Systems, pp.\  23--66. Springer
  Berlin Heidelberg.

\bibitem[\protect\citeauthoryear{Marshakova}{Marshakova}{1973}]{Marshakova1973system}
Marshakova, I.~V. (1973).
\newblock System of document connections based on references.
\newblock {\em Nauchno-Tekhnicheskaya Informatsiya Seriya 2 -- Informatsionnye
  Protsessy i Sistemy\/}~{\em 6}, 3--8.
\newblock (in Russian).

\bibitem[\protect\citeauthoryear{Merton}{Merton}{1957}]{merton1957social}
Merton, R.~K. (1957).
\newblock {\em {Social theory and social structure}}.
\newblock Free Press New York.

\bibitem[\protect\citeauthoryear{Milgram}{Milgram}{1967}]{milgram1967swp}
Milgram, S. (1967).
\newblock {The small world problem}.
\newblock {\em Psychology Today\/}~{\em 2\/}(1), 60--67.

\bibitem[\protect\citeauthoryear{Mitesser, Heinz, Havemann, and
  Gl{\"a}\-ser}{Mitesser et~al.}{2008}]{Mitesser2008mdr}
Mitesser, O., M.~Heinz, F.~Havemann, and J.~Gl{\"a}\-ser (2008).
\newblock {Measuring Diversity of Research by Extracting Latent Themes from
  Bipartite Networks of Papers and References}.
\newblock In H.~Kretschmer and F.~Havemann (Eds.), {\em Proceedings of WIS
  2008: Fourth International Conference on Webometrics, Informetrics and
  Scientometrics Ninth COLLNET Meeting}, Berlin. Humboldt-Universit{\"a}t zu
  Berlin: Gesellschaft f{\"u}r Wissenschaftsforschung.
\newblock ISBN: 978-3-934682-45-0, \url{http://www. collnet.
  de/Berlin-2008/MitesserWIS2008mdr. pdf}.

\bibitem[\protect\citeauthoryear{Moed, Gl{\"a}nzel, and Schmoch}{Moed
  et~al.}{2004}]{moed2004hqs}
Moed, H.~F., W.~Gl{\"a}nzel, and U.~Schmoch (Eds.) (2004).
\newblock {\em Handbook of Quantitative Science and Technology Research: The
  Use of Publication and Patent Statistics in Studies of S\&T Systems}.
\newblock Dordrecht: Kluwer Academic Publishers.

\bibitem[\protect\citeauthoryear{Morris, Yen, Wu, and Asnake}{Morris
  et~al.}{2003}]{morris2003tlv}
Morris, S.~A., G.~Yen, Z.~Wu, and B.~Asnake (2003).
\newblock {Time line visualization of research fronts}.
\newblock {\em Journal of the American Society for Information Science and
  Technology\/}~{\em 54\/}(5), 413--422.

\bibitem[\protect\citeauthoryear{Morris and Yen}{Morris and
  Yen}{2004}]{Morris2004crossmaps}
Morris, S.~A. and G.~G. Yen (2004).
\newblock Crossmaps: {Visualization} of overlapping relationships in
  collections of journal papers.
\newblock {\em Proceedings of the National Academy of Sciences of The United
  States of America\/}~{\em 101}, 5291--5296.

\bibitem[\protect\citeauthoryear{Mutschke, Mayr, Schaer, and Sure}{Mutschke
  et~al.}{2011}]{mutschke2011science}
Mutschke, P., P.~Mayr, P.~Schaer, and Y.~Sure (2011).
\newblock Science models as value-added services for scholarly information
  systems.
\newblock {\em Scientometrics\/}~{\em 89\/}(1), 349--364.

\bibitem[\protect\citeauthoryear{Newman}{Newman}{2001a}]{Newman2001scientific-I}
Newman, M. E.~J. (2001a).
\newblock Scientific collaboration networks. {I.} {Network} construction and
  fundamental results.
\newblock {\em Physical Review E\/}~{\em 64}, 016131.

\bibitem[\protect\citeauthoryear{Newman}{Newman}{2001b}]{Newman2001scientific-II}
Newman, M. E.~J. (2001b).
\newblock Scientific collaboration networks. {II.} {Shortest} paths, weighted
  networks, and centrality.
\newblock {\em Physical Review E\/}~{\em 64}, 016132.

\bibitem[\protect\citeauthoryear{Noyons}{Noyons}{2004}]{noyons2004science}
Noyons, E. C.~M. (2004).
\newblock {Science maps within a science policy context}.
\newblock In H.~F. Moed, W.~Gl{\"a}nzel, and U.~Schmoch (Eds.), {\em Handbook
  of Quantitative Science and Technology Research: The Use of Publication and
  Patent Statistics in Studies of S\&T Systems}, pp.\  237--256. Dordrecht:
  Kluwer Academic Publishers.

\bibitem[\protect\citeauthoryear{NRC}{NRC}{2005}]{nrc2005network}
NRC (2005).
\newblock Network science.
\newblock National Research Council of the National Academies, Washington, DC.

\bibitem[\protect\citeauthoryear{Ortega, Aguillo, Cothey, and
  Scharnhorst}{Ortega et~al.}{2008}]{ortega2008maps}
Ortega, J.~L., I.~Aguillo, V.~Cothey, and A.~Scharnhorst (2008).
\newblock {Maps of the academic web in the European Higher Education Area—an
  exploration of visual web indicators}.
\newblock {\em Scientometrics\/}~{\em 74\/}(2), 295--308.

\bibitem[\protect\citeauthoryear{Payette}{Payette}{2012}]{payette2012agent}
Payette, N. (2012).
\newblock Agent-based models of science.
\newblock In A.~Scharnhorst, K.~B\"orner, and P.~Besselaar (Eds.), {\em Models
  of Science Dynamics}, Understanding Complex Systems, pp.\  127--157. Springer
  Berlin Heidelberg.

\bibitem[\protect\citeauthoryear{Pinski and Narin}{Pinski and
  Narin}{1976}]{Pinski1976citation}
Pinski, G. and F.~Narin (1976).
\newblock Citation influence for journal aggregates of scientific publications:
  theory, with application to literature of physics.
\newblock {\em Information Processing \& Management\/}~{\em 12}, 297--312.

\bibitem[\protect\citeauthoryear{Prabowo, Thelwall, Hellsten, and
  Scharnhorst}{Prabowo et~al.}{2008}]{prabowo2008evolving}
Prabowo, R., M.~Thelwall, I.~Hellsten, and A.~Scharnhorst (2008).
\newblock {Evolving debates in online communication -- a graph analytical
  approach}.
\newblock {\em Internet Research\/}~{\em 18\/}(5), 520--540.

\bibitem[\protect\citeauthoryear{Pyka and Scharnhorst}{Pyka and
  Scharnhorst}{2009}]{pyka2009intro}
Pyka, A. and A.~Scharnhorst (2009).
\newblock {Introduction. Network perspectives on innovations: Innovative
  networks -- network innovation}.
\newblock In A.~Pyka and A.~Scharnhorst (Eds.), {\em {Innovation Networks --
  New Approaches in Modelling and Analyzing}}. Berlin: Springer.

\bibitem[\protect\citeauthoryear{Rip and Courtial}{Rip and
  Courtial}{1984}]{rip1984co-word}
Rip, A. and J.~Courtial (1984).
\newblock {Co-word maps of biotechnology: An example of cognitive
  scientometrics}.
\newblock {\em Scientometrics\/}~{\em 6\/}(6), 381--400.

\bibitem[\protect\citeauthoryear{Salton and McGill}{Salton and
  McGill}{1983}]{Salton1983imi}
Salton, G. and M.~J. McGill (1983).
\newblock {\em Introduction to Modern Information Retrieval}.
\newblock New York, NY, USA: McGraw-Hill, Inc.

\bibitem[\protect\citeauthoryear{Scharnhorst}{Scharnhorst}{2001}]{scharnhorst2001constructing}
Scharnhorst, A. (2001).
\newblock {Constructing Knowledge Landscapes within the Framework of
  Geometrically Oriented Evolutionary Theories}.
\newblock In M.~Matthies, H.~Malchow, and J.~Kriz (Eds.), {\em Integrative
  Systems Approaches to Natural and Social Sciences -- Systems Science 2000},
  pp.\  505--515. Berlin: Springer.

\bibitem[\protect\citeauthoryear{Scharnhorst}{Scharnhorst}{2003}]{scharnhorst2003complex}
Scharnhorst, A. (2003, July).
\newblock Complex networks and the web: Insights from nonlinear physics.
\newblock {\em Journal of Computer-Mediated Communication\/}~{\em 8\/}(4).
\newblock \url{http://jcmc.indiana.edu/vol8/issue4/scharnhorst.html}.

\bibitem[\protect\citeauthoryear{Skupin}{Skupin}{2009}]{skupin2009discrete}
Skupin, A. (2009).
\newblock {Discrete and continuous conceptualizations of science: Implications
  for knowledge domain visualization}.
\newblock {\em Journal of Informetrics\/}~{\em 3\/}(3), 233--245.
\newblock Special Edition on the Science of Science: Conceptualizations and
  Models of Science, ed. by Katy B{\"o}rner \& Andrea Scharnhorst.

\bibitem[\protect\citeauthoryear{Small}{Small}{1973}]{Small1973cocitation}
Small, H. (1973).
\newblock {Co-citation in the Scientific Literature: A New Measure of the
  Rela\-tionship Between Two Documents}.
\newblock {\em Journal of the American Society for Information Science\/}~{\em
  24}, 265--269.

\bibitem[\protect\citeauthoryear{Small}{Small}{1978}]{small1978cited}
Small, H. (1978).
\newblock Cited documents as concept symbols.
\newblock {\em Social studies of science\/}~{\em 8\/}(3), 327.

\bibitem[\protect\citeauthoryear{Small and Sweeney}{Small and
  Sweeney}{1985}]{small1985ctc}
Small, H. and E.~Sweeney (1985).
\newblock {Clustering the Science Citation index using co-citations}.
\newblock {\em Scientometrics\/}~{\em 7\/}(3), 391--409.

\bibitem[\protect\citeauthoryear{Snijders, {van de Bunt}, and
  Steglich}{Snijders et~al.}{2010}]{snijders2010introduction}
Snijders, T. A.~B., G.~G. {van de Bunt}, and C.~E.~G. Steglich (2010).
\newblock Introduction to stochastic actor-based models for network dynamics.
\newblock {\em Social Networks\/}~{\em 32\/}(1), 44--60.

\bibitem[\protect\citeauthoryear{Sonnenwald}{Sonnenwald}{2007}]{sonnenwald2007scientific}
Sonnenwald, D. (2007).
\newblock {Scientific collaboration}.
\newblock {\em Annual Review of Information Science and Technology\/}~{\em
  41\/}(1).

\bibitem[\protect\citeauthoryear{Swanson}{Swanson}{1986}]{swanson1986fish}
Swanson, D.~R. (1986).
\newblock {Fish oil, Raynauds syndrome, and undiscovered public knowledge}.
\newblock {\em Perspectives in Biology and Medicine\/}~{\em 30\/}(1), 7--18.

\bibitem[\protect\citeauthoryear{Thelwall}{Thelwall}{2004}]{thelwall2004link}
Thelwall, M. (2004).
\newblock {\em {Link analysis: An information science approach}}.
\newblock Academic Press.

\bibitem[\protect\citeauthoryear{Thelwall}{Thelwall}{2009}]{thelwall2009introduction}
Thelwall, M. (2009).
\newblock {Introduction to Webometrics: Quantitative Web Research for the
  Social Sciences}.
\newblock In {\em Synthesis Lectures on Information Concepts, Retrieval, and
  Services}, Volume~1, pp.\  1--116. Morgan \& Claypool Publishers.

\bibitem[\protect\citeauthoryear{{van den Heuvel}}{{van den
  Heuvel}}{2008}]{vandenHeuvel2008building}
{van den Heuvel}, C. (2008).
\newblock {Building Society, Constructing Knowledge, Weaving the Web. Otlet's
  visualizations of a global information society and his concept of a universal
  civilization}.
\newblock In W.~B. Rayward (Ed.), {\em European Modernism and the Information
  Society}, Chapter~7, pp.\  127--153. London: Ashgate Publishers.

\bibitem[\protect\citeauthoryear{van~den Heuvel and Rayward}{van~den Heuvel and
  Rayward}{2011}]{vanHeuvel2011facing}
van~den Heuvel, C. and W.~Rayward (2011).
\newblock {Facing interfaces: Paul Otlet's visualizations of data integration}.
\newblock {\em Journal of the American Society for Information Science and
  Technology\/}~{\em 62\/}(12), 2313--2326.

\bibitem[\protect\citeauthoryear{{van den Heuvel} and Rayward}{{van den Heuvel}
  and Rayward}{2005}]{heuvel2005visualizing}
{van den Heuvel}, C. and W.~B. Rayward (2005, July).
\newblock {Visualizing the Organization and Dissemination of Knowledge: Paul
  Otlet's Sketches in the Mundaneum, Mons', Envisioning a Path to the Future}.
\newblock Weblog.
\newblock
  \url{http://informationvisualization.typepad.com/sigvis/2005/07/visualizations_.html}.

\bibitem[\protect\citeauthoryear{{van der Eijk}, {van Mulligen}, Kors, Mons,
  and {van den Berg}}{{van der Eijk} et~al.}{2004}]{vanderEijk2004constructing}
{van der Eijk}, C.~C., E.~M. {van Mulligen}, J.~A. Kors, B.~Mons, and J.~{van
  den Berg} (2004).
\newblock {Constructing an Associative Concept Space for Literature-Based
  Discovery}.
\newblock {\em Journal of the American Society for Information Science and
  Technology\/}~{\em 55\/}(5), 436--444.

\bibitem[\protect\citeauthoryear{Vlach{\`y}}{Vlach{\`y}}{1978}]{vlachy1978field}
Vlach{\`y}, J. (1978).
\newblock {Field mobility in Czech physics-related institutes and faculties}.
\newblock {\em Czechoslovak Journal of Physics\/}~{\em 28\/}(2), 237--240.

\bibitem[\protect\citeauthoryear{Wagner}{Wagner}{2008}]{wagner2008new}
Wagner, C.~S. (2008).
\newblock {The new invisible college}.
\newblock Science for Development, Brookings Institution, Washington, DC.

\bibitem[\protect\citeauthoryear{Wasserman and Faust}{Wasserman and
  Faust}{1994}]{Wasserman1994social}
Wasserman, S. and K.~Faust (1994).
\newblock {\em Social Network Analysis: Methods and Applications}.
\newblock Cambridge University Press.

\bibitem[\protect\citeauthoryear{Wouters}{Wouters}{1999}]{wouters1999citation}
Wouters, P. (1999).
\newblock {The citation culture}.
\newblock {\em Unpublished Ph. D. Thesis, University of Amsterdam\/}.
\newblock \url{http://garfield.library.upenn.edu/wouters/wouters.pdf}.

\end{thebibliography}
\bibliographystyle{chicago}

\end{document}